\documentclass[aps,prl,twocolumn,superscriptaddress,balancelastpage,showpacs,reprint]{revtex4-1}
\usepackage{graphicx}
\usepackage{amsmath}
\usepackage{dcolumn}
\usepackage{bm}
\usepackage{mathtools}
\usepackage{color,soul}

\begin{document}

\title{Mutation at Expanding Front of Self-Replicating Colloidal Clusters}

\author{Hidenori Tanaka}
\email{tanaka@g.harvard.edu}
\affiliation{Harvard John A. Paulson School of Engineering and Applied Sciences, Harvard University, Cambridge, MA 02138}
\affiliation{Kavli Institute for Bionano Science and Technology, Harvard University, Cambridge, MA 02138} 

\author{Zorana Zeravcic}
\affiliation{Harvard John A. Paulson School of Engineering and Applied Sciences, Harvard University, Cambridge, MA 02138}
\affiliation{Soft matter and chemistry laboratory, ESPCI PSL Research University, 75005 Paris, France}

\author{Michael P. Brenner}
\affiliation{Harvard John A. Paulson School of Engineering and Applied Sciences, Harvard University, Cambridge, MA 02138}
\affiliation{Kavli Institute for Bionano Science and Technology, Harvard University, Cambridge, MA 02138}

\date{\today}
\pacs{
81.05.Zx, 
81.16.Dn, 
87.23.Cc, 
82.70.Dd  
}

\begin{abstract}
We construct a scheme for self-replicating square clusters of particles in two spatial dimensions, and validate it with computer simulations in a finite-temperature heat bath. We find that the self-replication reactions propagate through the bath in the form of Fisher waves. Our model reflects existing colloidal systems, but is simple enough to allow simulation of many generations and thereby the first study of evolutionary dynamics in an artificial system. By introducing spatially localized mutations in the replication rules, we show that the mutated cluster population can survive and spread with the expanding front in circular sectors of the colony.
\end{abstract}

\maketitle

Self-replication followed by mutation and evolution is a key driver of biological complexity. The change of fitness landscapes due to environmental conditions creates the driving force for the evolution of new functionalities. A holy grail of modern materials science research is to emulate this natural evolution of functionality, and to design materials systems where evolution based discovery strategies could apply. Creating a population dynamics in materials requires both designing efficient self-replication, as well as some mechanism for mutating the dynamics so that novel structures can arise. 

Recently, initial steps towards creating self-replicating artificial materials have been taken using DNA nanotechnology \cite{Leunissen:2009ui,Wang:2011vx,Schulman:2012vo,2015NatNa..10..528K},
light switchable colloidal dimers \cite{Light-swichable} or magnetic dipolar colloids \cite{PhysRevE.92.042305}.  With DNA based interactions, hybridization causes specific, short range interactions between nano/micro scale components, and both the specificity \cite{Feng:2013dp, Leunissen:2009p7109} and the timescales \cite{Turberfield:2000ex} of the interactions between strands can be chosen by tuning the DNA sequences. A striking study \cite{Wang:2011vx} uses chains of DNA tiles as templates for replication, achieving a few generations of replication in experiments. More recently, these ideas were extended to ring structures \cite{2015NatNa..10..528K}  of DNA tile motifs. A theoretical study \cite{2014PNAS..111.1748Z} demonstrated how specific interactions can lead to self-replication of finite clusters of particles.  

Up until now, there has been no explicit demonstration of  mutation/amplification cycles  in a self-replicating material, either theoretically or experimentally. A major challenge has been the difficulty in either experiments or simulations of producing enough replication cycles that a meaningful mutation could be carried out. In contrast, selected evolution of DNA or RNA is a common technology due to the highly efficient and optimized Polymerase Chain Reaction \cite{innis2012pcr}. 
In this paper, we introduce a model of self-replicating clusters that is computationally tractable enough that we can explicitly study the emergence of mutations. We demonstrate a phenomenology that is strikingly similar to the development of mutations in growing bacterial colonies \cite{Nelson:Bacteria,korolev2010genetic,Korolev:2011iy}.

Our  system consists of colloids coated with specific DNA strands, in which motion of colloids are confined in two dimensions. The self-replicating objects are colloidal clusters that make their progeny through a geometrical templating scheme introduced earlier \cite{2014PNAS..111.1748Z}. The two dimensional system that  we model is directly related to recent experiments on clusters of identical colloidal particles \cite{perry2015two}.  To allow the complexity of self-replication and mutation/selection phenomena, we model colloids with specific interactions, such as ones already realized in self-assembly experiments \cite{Jones:2015vc,Wu:2013jg,Rogers:2015bv,Rogers:2011uz,2012Natur.491...51W,schade2013tetrahedral,Brodin:2015dq}. The fact that the system is two dimensional gives substantial computational savings and makes it possible to simulate large systems with $>20$ generations (see SI section I for the definition) of progeny. 
 
We find two striking features of our model of self-replicating clusters:
First, when the clusters grow exponentially, they deplete the monomer pool from the bath, and cause the formation of a propagating front obeying Fisher's dynamics \cite{FISHER:1937hy,kolmogorov}, that expands into the monomer bath. 
Second, a single cluster with a changed replication rule (``mutation'') can initiate a population change within the expanding front and form a sector of mutated structures (``genetic drift''). As we will argue below, we believe our model presents the first observation of evolutionary dynamics in a controlled artificial self-replicating system.

\begin{figure}[!tbp]
\begin{center}
\includegraphics[width=\columnwidth]{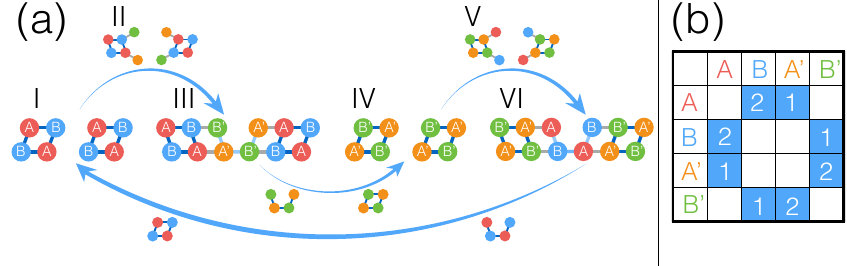}
\caption{ (a) Self-replication scheme of square clusters with $n_{\rm{c}}=3$. The reaction starts with two parent square clusters composed of A and B particle species, having permanent bonds (stage I). Complementary monomers (A',B') attach to the parent clusters (stage II) and attraction among the attached monomers leads to templating of another 4-mer structure (stage III). The thus formed complementary cluster templates a new cluster in the same way, thereby closing the hypercycle (stages IV to VI).
(b) Interaction matrix of particle species. Blue matrix entries represent attractive interaction, and white entries represent repulsive force, while the inscribed number specifies the species valence.}
\label{figure1}
\end{center}
\end{figure}

In our system, the surface of each colloid is covered by specific stickers (i.e., DNA strands) which mediate specific short range attraction between pairs of particles. We choose to consider replication of square clusters, as a simple example that is not constrained to a linear geometry.
When each particle in the parent cluster can attach at most one complementary monomer, geometrical constraints prevent direct templating from a single parent. Inspired by previous work \cite{2014PNAS..111.1748Z}, we circumvent this problem by templating a square cluster between two parent square clusters, Fig.~\ref{figure1}(a).
The self-replication scheme begins when two parent square clusters, each composed of A, B particle species (stage I), each attach complementary particles, species A', B', respectively, from the monomer bath (see Fig.~\ref{figure1}(b)), thereby forming two independent hairy squares (stage II). Then, attraction between A' and B' allows bonding of the two hairy clusters (stage III). The formed structure of complementary particles is detached from the parents (stage IV), forming a square. The square made of A',B' particles now becomes a parent involved in templating its complementary (A,B) square thus closing the hypercycle (stages V, VI).

The detachment step between stages III and IV (Fig.~\ref{figure1}(a)) is critical in self-replication reactions
\footnote{In experiments, detachment requires energy input and is facilitated by temperature and UV-light cycling \cite{Wang:2011vx}, while several other possibilities have been proposed \cite{PhysRevE.92.042305,Zhang:2013bz}.}.
In our simulations, we model the detachment process as in Ref.\cite{2014PNAS..111.1748Z}: 
When the number of bonds $n$ between attached monomers (while they are attached to their respective parents) reaches the critical value $n_{\rm{c}}$, the bonds between parent particles and monomers are removed and the existing bonds between the monomers become irreversible (see SI section II for details). In practice, $n$ can increase by more than one within a single simulation timestep. Therefore, for a chosen $n_{\rm{c}}$, a newly formed structure can have $n\geq n_{\rm{c}}$ and a non-square geometry, and can become a parent for future reactions that can significantly deviate from the scheme in Fig.~\ref{figure1}(a).

\begin{figure}[!tbp]
\begin{center}
\includegraphics[width=\columnwidth]{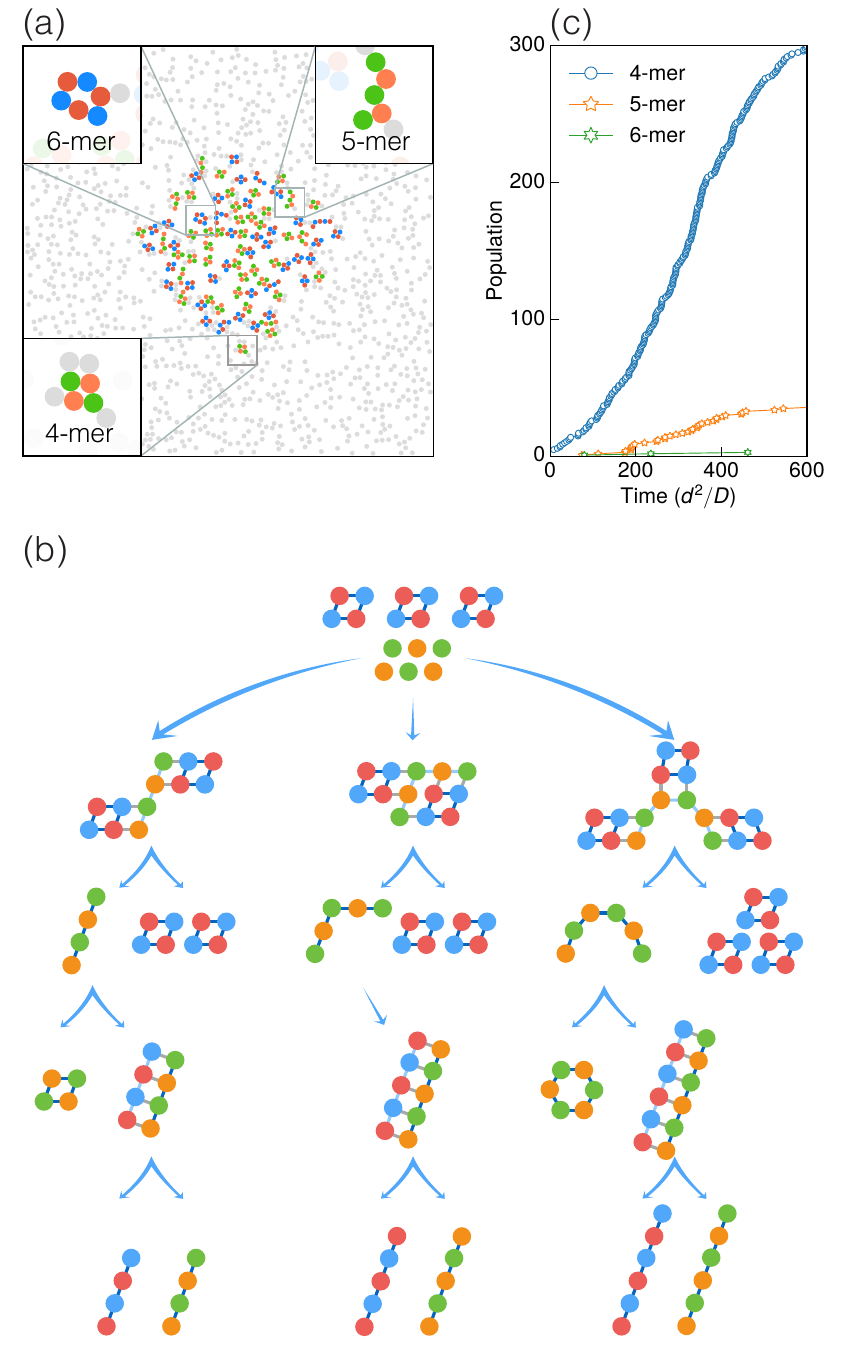}
\caption{Brownian dynamics simulation of self-replicating colloidal clusters using 1652 colloidal particles.  (a) A snapshot from our simulation: Particles are colored according to their species (Fig.~\ref{figure1}(a)) if they comprise a cluster; otherwise they are gray. This snapshot shows successful replication of desired 4-mer (square) cluster as well as other undesired 5-mer and 6-mer structures. (b) Example of frequent replication pathways leading to formation of each of the 4-, 5- and 6-mers. (c) Population of 4-mers, 5-mers and 6-mers as a function of time.
}
\label{figure2}
\end{center}
\end{figure}
\begin{figure*}
\begin {center}
\centerline{\includegraphics[width = \textwidth]{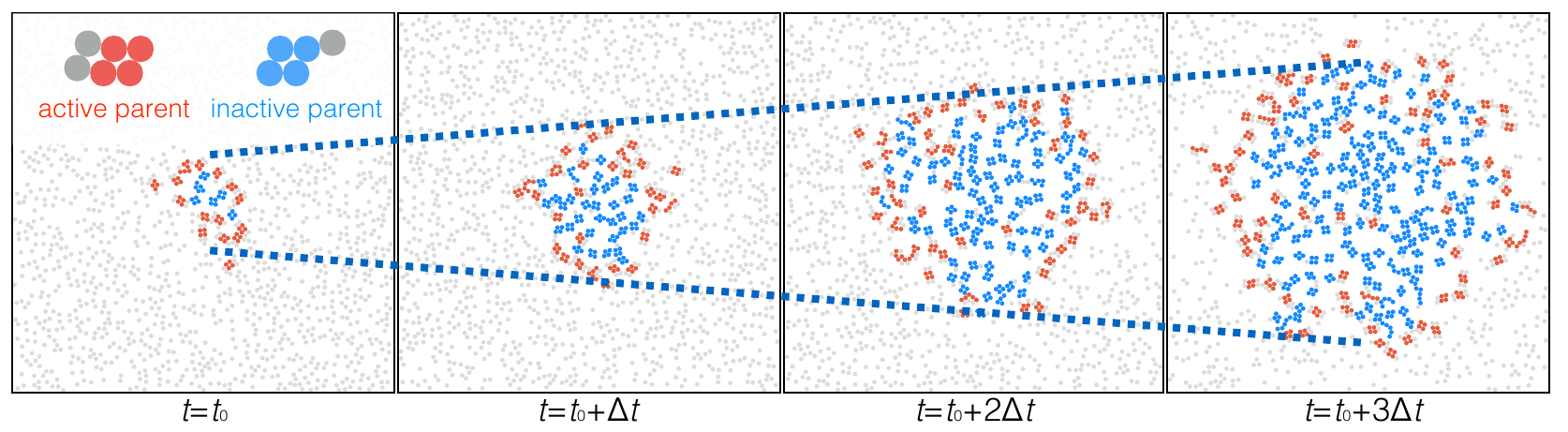}}
\caption{Simulation snapshots taken at $\Delta t=100d^2/D$ intervals. Parent clusters with at least two attached monomers are colored red (active parents) and parent clusters with less than two attached monomers are colored blue (inactive parents). Dashed lines ($\propto t$) are guides to the eye.}
\label{figure3}
\end{center}
\end{figure*}

To investigate emergent behavior in this system, we perform Brownian dynamics simulation with periodic boundary condition. The dynamics of the $i$-th particle is governed by the over-damped Langevin equation 
\begin{equation}
\partial_{t} \bm{ r}_{i} = -\mu \sum_{j\neq i} \bm{ \nabla } U(r_{ij}) + \bm{ \eta }_i (t),
\end{equation}
where $\bm{ r}_{i}$ is the position vector of $i$-th particle, $r_{ij}\equiv |\bm{r}_{i}-\bm{ r}_{j}|$, and $\mu$ is the mobility. The short range interaction between $i$-th and $j$-th particle $U(r_{ij})$ is given by a modified Morse potential in case of attraction, and harmonic potential in case of repulsion (see SI section III for details). The interaction range is $1.05d$ in case of attraction and $1.0d$ in case of repulsion, where $d$ is the particle diameter. The implementation of short ranged interactions is validated against recent experimental results on colloidal clusters in two spatial dimensions \cite{perry2015two}  (see SI section IV).
The noise term $\bm{ \eta }_i (t)$ satisfies the fluctuation-dissipation relation $\big \langle\eta_{i\alpha}(t)\eta_{j\beta}(t') \big \rangle= 2D\delta_{ij} \delta_{\alpha \beta} \delta(t-t') $ where $\alpha,\beta=x,y$, and $D$ is the diffusion constant of monomers. In our simulation, we measure length in units of particle diameter $d$, time in $d^2/D$, energy in $k_{\rm{B}}T$.

We simulate a system of $N=1652$ particles with an area fraction $\phi=0.193$ in a square box ($L=82d$) with periodic boundary conditions, and we keep the temperature fixed. Sixteen out of $N$ particles comprise four initial parent square clusters placed around the center.
We set $n_c=3$, so that detachment reaction occurs when at least three bonds form between attached monomers. To simplify simulations, we do not allow the free monomers to interact with each other.
Fig.~\ref{figure2}(a) shows a typical snapshot from our simulation. 
As clusters replicate and diffuse, they form a circular colony. The colony mainly contains our designed structure, the square, but we also find geometrically distinct structures of 4, 5 or 6 particles, whose formation is allowed by the detachment criterion $n_c=3$. Since there is no attraction between particles of the same species (Fig.~\ref{figure1}(b)), 5-mers can only exist as chains while 4- and 6-mers only form squares and hexagons, respectively (see insets of panel Fig.~\ref{figure2} (a)).
Examples of frequently observed replication pathways of these structures are shown in Fig.~\ref{figure2}(b) (see SI section V for more details).
Fig.~\ref{figure2}(c) shows the time dependence of the population of each of these structures, demonstrating that with $n_c=3$ the square clusters dominate.

Using our observation of self-replication over generations we analyze properties of spatiotemporal structure of the colony. Fig.~\ref{figure3} illustrates the spatial distribution of replication reactions.
We find that a good indicator \footnote{After trying several criteria, we find that the great majority of parent clusters with two or more attached monomers in a snapshot do create daughters within a short time window after that snapshot. Conversely, we find that the great majority of daughters are created by ``active'' parents. By labeling parents with three or more attached monomers ``active'' is also a good indicator.} of where the replication happens is the location of parents with at least two attached monomers, and we label such parents ``active'' (colored red). Most of the active parents localize in a thin band at the boundary of the colony, whereas most of the clusters in the interior are ``inactive'' (colored blue).

\begin{figure}[!tbp]
\begin{center}
\includegraphics[width=\columnwidth]{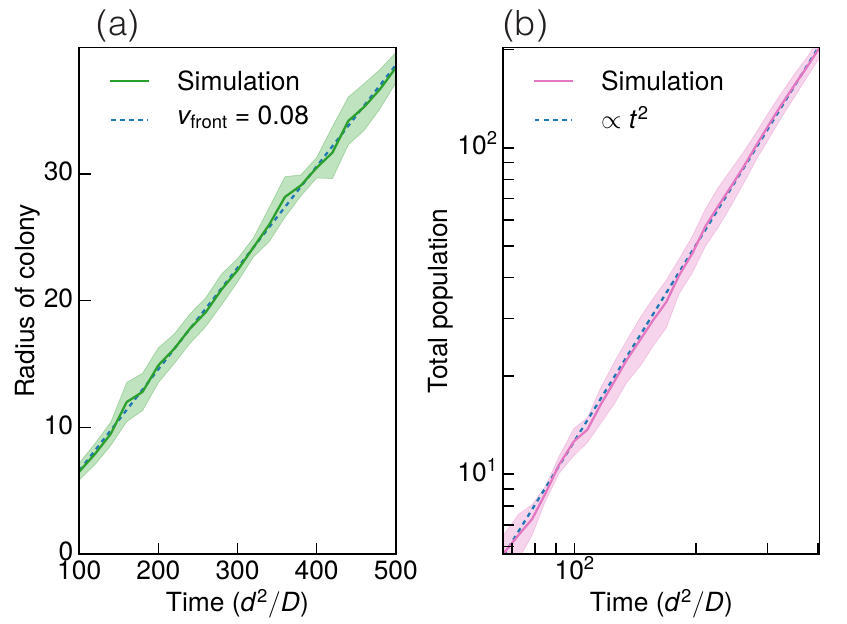}
\caption{(a) Log-log plot of the total population as a function of time. The solid line is the average over 10 independent simulations. The shaded region shows one standard deviation above and below the average, while the dashed line ($\propto t^2$) is the best linear fit. (b) Radius of the circular colony as function of time. The solid line is the average over 10 independent simulations. The shaded region shows one standard deviation above and below the average. The dashed line is the best linear fit, suggesting emergence of expanding front with constant velocity.}  
\label{figure4}
\end{center}
\end{figure}

To understand the spatial spreading of the colony, we consider the time evolution of a coarse-grained population density field of clusters $u(\bm{r},t)$ through a Fisher-Kolmogorov-Petrovskii-Piskunov (F-KPP) type of reaction-diffusion equation \cite{FISHER:1937hy,kolmogorov}:
\begin{equation}
\frac{\partial u}{\partial t} =  \alpha u (1-u ) + D_{\rm{eff}} \nabla^2 u ,
\end{equation}
with initial growth rate $\alpha$, and an effective diffusion constant $D_{\rm{eff}}$. This equation has an asymptotic traveling wave solution of circular symmetry $u(r,t)=f(r-v_{\rm{front}}t)$, where the front velocity
\begin{equation}
  \label{eq:3}
v_{\rm{front}} =2\sqrt{D_{\rm{eff}}\alpha}  
\end{equation}
at large $r$ \cite{van2003front,murray2002mathematical}. Fig.~{\ref{figure4}}(a) shows that the colony radius moves at constant velocity, measured to be $v_{\rm{front}}=0.08D/d$. We also directly measure the initial growth rate of the colony to be $\alpha=1.5 \times 10^{-2}D/d^2$, giving the estimate $D_{\rm{eff}}\approx0.11D$ from Eq.~\eqref{eq:3}. To validate the F-KPP dynamics entailed by Eq.~\eqref{eq:3}, from our simulations we directly measure the effective diffusion constant of all clusters, $D^{\rm{sim}}_{\rm{eff}}$, which takes into account that clusters are found in different stages of self-replication reaction, Fig.~\ref{figure1} (a) (see SI section VI for details).
We find $D^{\rm{sim}}_{\rm{eff}}=0.12\pm 0.04D$, in good agreement with the estimated $D_{\rm{eff}}$.

The F-KPP dynamics implies a length scale
\begin{equation}
  \label{eq:2}
  \lambda=\sqrt{D_{\rm{eff}}/\alpha}
\end{equation}
which sets the width of the traveling front (Fisher wave), estimated here to be $\lambda= 3d$. When the system size $L$ is much less than $\lambda$, one can only observe an exponential growth of the colony without formation of a front. In the opposite regime $\lambda\ll L$ one observes an initial exponential growth followed by formation of a propagating front and growth of colony as a power-law $t^d$ with $d$ the spatial dimension. Our simulations are in the latter regime, exhibiting the propagation of fully-formed front whose estimated width of order $\lambda=3d$ is consistent with the width of region populated by active parents in Fig.~\ref{figure3} (see SI section VII). In Fig.~\ref{figure4}(b) we show the colony size, i.e., the total population of all clusters over time, averaged over 10 independent simulations. The growth curve is quadratic in time (after initial exponential expansion) as expected from F-KPP dynamics. An intuitive explanation of this power-law follows from assuming that the replication process occurs within the front which has a constant width. Then the number of newly created clusters per unit time at time $t$ is proportional to $\sqrt{N_{\rm{clust}}(t)}$, which leads to the quadratic population growth.

\begin{figure}[!tbp]
\begin{center}
\includegraphics[width=\columnwidth]{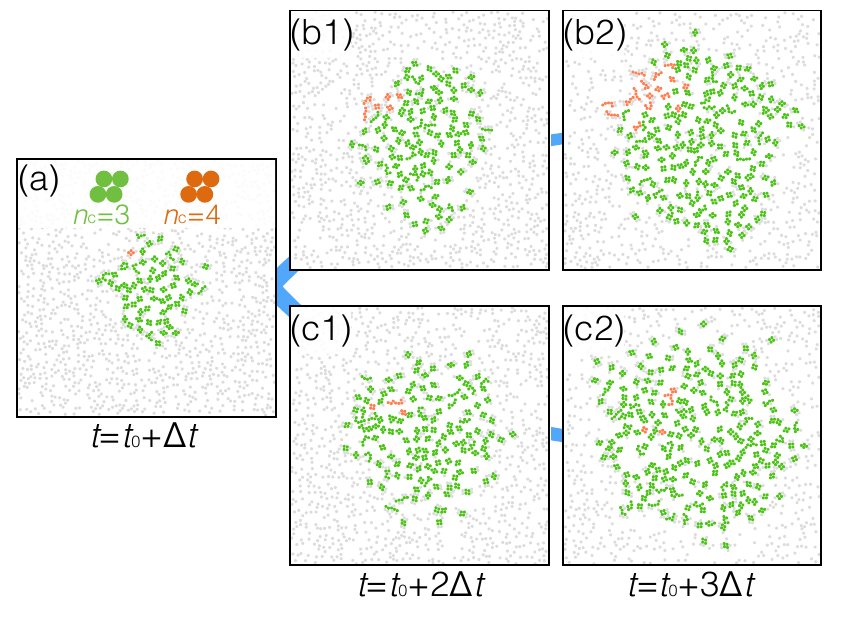}
\caption{(a) The second snapshot ($t=t_{0}+\Delta t$) of Fig. \ref{figure3} is shown, where one square cluster (orange) at the boundary is mutated by assigning it $n_{\rm{c}}=4$. All other clusters (green) retain $n_{\rm{c}}=3$. 
(b1-b2) and (c1-c2) are results from two independent simulations started from (a). The mutated progeny either dominates in a circular sector of the colony (b1) and survives with expanding front (b2); or stays within the colony bulk (c1) and stops reproducing (c2).
} 
\label{figure5}
\end{center}
\end{figure}

Finally, we consider mutations in the cluster population.
We seek to define a ``mutation'' as an hereditary trait that leads to altered properties.
Within our model, the size and shape of a parent will not persist through many generations of progeny (see Fig.~\ref{figure2}(b)).
We can however define a mutation as a change of the replication rule $n_{\rm{c}}=3$ into $n_{\rm{c}}=4$: First, this property can be inherited by every daughter of the mutated cluster (a dominant trait); second, a mutated population has a strikingly different distribution of cluster sizes than the non-mutated population. In particular, $n_{\rm{c}}=3$ population is majority 4-mers with only $\sim 10\%$ of 5-mers (see Fig.~\ref{figure2}(c) and S2(a)), while mutated population with $n_{\rm{c}}=4$ has $\sim 70\%$ of 5-mers (see Fig.~S2(b)). 
To investigate mutations, we start from the colony in the second panel of Fig.~\ref{figure3}, and select a single square cluster at the edge of colony, Fig.~\ref{figure5}(a). We mutate this cluster, leaving all others intact.
Fig.~\ref{figure5} shows two different outcomes of 13 simulations started with this initial condition: In panels (b1-b2), the population of mutated structures grows at the expanding front and dominates in a circular sector. As expected, the population in the sector is majority 5-mers. In contrast, in panels (c1-c2) the mutation goes extinct, since the progeny of the mutated cluster stops reproducing in the monomer-deficient colony bulk.

To judge how mutation affects the ``fitness'' of population, in each simulation we trace the progeny of the mutated cluster and the progeny of several randomly selected non-mutated clusters at the opposite side of the colony front. The outcomes indicate that the survival probability for the mutated lineage is lower $(62\%\pm15\%)$ than for the non-mutated $(83\%\pm11\%)$ (for details see SI section VIII).
We relate the lowered fitness of mutated population to the fact that the mutated clusters on average need to consume more monomers compared to their non-mutated competitors on the front. The competitive advantage can be observed by comparing a purely non-mutated and a purely mutated colony: The rate of monomer consumption is similar (Fig.~S6(b)), while the growth rate of the mutated colony lags behind (Fig.~S6(c)) as described in SI section VIII.
The fact that survival of a (non-)mutated cluster lineage is subject to randomness by definition entails ``genetic drift''. In our system we directly observe the genetic drift because lineages propagate spatially with the Fisher's wave.

In summary, we have introduced a model of self-replicating colloidal clusters which despite its simplicity shows remarkably rich behavior including Fisher wave propagation, and the possibility of studying mutations, fitness and genetic drift, as some of key components of evolutionary dynamics. Very recently, propagating reaction-diffusion fronts have been observed in different synthetic self-replicating systems at the molecular scale \cite{Zadorin:2015hj,Bottero:2016by}. Our system relates to an artificial material at the mesoscale. We remark that the basic ingredients required to experimentally realize our system are becoming available. Controlled valence of isotropic mesoscale particles has been demonstrated \cite{Feng:2013dp,angioletti2014mobile}. We believe that the detachment step could be realized with time-dependent interactions that can either strengthen (among attached monomers) or weaken (between the parent cluster and attached monomers) in time \cite{Anonymous:gJOl57ns}, which require consumption of energy. A first step towards time-dependent interactions has been realized between nanoparticles using complex strand-displacement reactions that rely on a DNA fuel source \cite{Yao:2015cz,Turberfield:2000ex}. An alternative would be to globally cycle temperature and light to achieve the detachment process and cluster stabilization, a strategy already used in some studies \cite{Leunissen:2009ui,Wang:2011vx}. The final ingredient --- implementation of the mutation mechanism --- is still an open problem. We require a change in the characteristic time of the detachment process, therefore changing which structure is preferred by the replication process. At the same time, this change in the replication process needs to be inherited by daughters. Although a specific mechanism for this is unclear at the moment, we believe the flexibility provided by the DNA nanotechnology is sufficient to realize it. Our concept is, however, applicable to artificial systems at different scales and opens a new door for implementing evolutionary dynamics in experiments.

\bigskip

\textbf{ Acknowledgement }
H.T. would like to thank H.~Isobe, K.~Kawaguchi, A.A.~Lee and members of Manoharan lab for fruitful discussions. 
H.T. is supported by the Funai Foundation for Information Technology through Funai Overseas Scholarship.
The computations in this paper were run on the Odyssey cluster supported by the FAS Division of Science, Research Computing Group at Harvard University.
This research was funded by the National Science Foundation through Grant DMR-1435964, the Harvard Materials Research Science and Engineering Center Grant DMR1420570, and the Division of Mathematical Sciences Grant DMS-1411694. M.P.B. is an investigator of the Simons Foundation.
\bibliography{lib}
\end{document}


\title{Supplemental Material\\
Mutation at Expanding Front of Self-Replicating Colloidal Clusters}
\author{Hidenori Tanaka}
\affiliation{Harvard John A. Paulson School of Engineering and Applied Sciences, Harvard University, Cambridge, MA 02138}
\affiliation{Kavli Institute for Bionano Science and Technology, Harvard University, Cambridge, MA 02138}
\email{tanaka@g.harvard.edu}
\author{Zorana Zeravcic}
\affiliation{Soft matter and chemistry laboratory, ESPCI, 75005 Paris, France}
\affiliation{Harvard John A. Paulson School of Engineering and Applied Sciences, Harvard University, Cambridge, MA 02138}
\author{Michael P. Brenner}
\affiliation{Harvard John A. Paulson School of Engineering and Applied Sciences, Harvard University, Cambridge, MA 02138}
\affiliation{Kavli Institute for Bionano Science and Technology, Harvard University, Cambridge, MA 02138}

\date{\today}

\maketitle

\section{I. Tracing Generations}
In our system, parent clusters can mate across generations to template a daughter cluster. Therefore, the family tree forms a network rather than a simple tree structure. We define the generation of a daughter cluster by incrementing by one the largest generation among her parents. For example, when the parent clusters of the 2nd and the 7th generations template a new daughter, we assign it to be the 8th generation. According to this definition, we observed self-replication for more than 20 generations in our simulations.

\section{II. Implementation of self-replication mechanism}
We start our simulation with 4 square clusters, two composed of A-B species and two of the complementary A'-B' species. The rest of the particles are free monomers that we don't assign any species in the beginning. Only once a monomer comes into contact with a cluster particle (inter-particle distance $\leq 1.05d$) we assign it the complementary species. This simplification allows us to accelerate our simulation. At each time step, for each cluster, we: (i) identify all monomers attached to the clusters, (ii) identify all the contiguous networks which these monomers form through contacts amongst each other, (iii) count the number of bonds $n$ amongst monomers in each of their contiguous networks and (iv) if $n\geq n_{\rm{c}}$ we initiate  ``detachment'' for that network. The detachment process for a group of monomers consists of defining them as a new cluster, making the bonds amongst them irreversible (see SI section III), and making their interactions with all other clusters repulsive.

\section{III. Particle interactions}

The interaction potential between two colloidal particles $i$ and $j$ is modeled by combining the harmonic potential for the core repulsion part and the Morse potential for the DNA mediated short range attraction part:
\begin{equation}
U_{0}(r_{ij}) = \\
\begin{cases}
  \frac{1}{2}k( r_{ij} - d )^2 -E_{ij} & \text{($r_{ij} \leq d$)} \\
  E_{ij} e^{- \rho (r_{ij}-1) } (e^{-\rho(r_{ij}-1)} - 2 )  & \text{($ d  <  r_{ij}   \leq r_{\rm{c}}$)  }\\
  0 & \text{($r_{\rm{c}}<r_{ij}$)}
\end{cases}
\end{equation}
where $U_{0}$ is the untruncated form of the potential, $r_{ij}$ is the distance between the two particles, $k$ is the spring constant, $d$ is the diameter of a particle, $E_{ij}$ is the depth of potential which is specified by the interaction matrix, and $\rho$ is the range parameter of the Morse potential. To represent core repulsion, the spring constant is set to be $k=10^5(k_{\rm{B}}T/d^2)$ which constrains the time step to be $2\times10^{-6} (d^2/D)$.
We set the range parameter of the Morse potential $\rho=80$ and smoothly truncate entire potential at $r_{\rm{c}}=1.05d$ by adding a linear term as below
\begin{equation}
U(r_{ij})=U_{0}(r_{ij})-(U_{0}(r_{\rm{c}}) + U_{0}'(r_{\rm{c}})(r_{ij}-r_{\rm{c}})),
\end{equation}
which is the final form of the interaction potential used in our simulations. In the experiments the DNA mediated attraction has the interaction range of $\sim 2\%$ of particle diameter \cite{Rogers:2011uz}, whereas we choose $5\%$ to ensure numerical stability of the simulation. A similar interaction potential has been used to study free-energy landscapes of colloidal cluster with short-ranged interaction potentials \cite{holmes2013geometrical}.

In our simulations the bond strength is set by the depth of the truncated potential $E_{ij}$. When the particles are two attached monomers (A'-B' or A-B bond), we choose $E_{ij}=E\equiv 10k_{\rm{B}}T$. This value allows the bonds between attached monomers to be broken by thermal noise. We set the bond strength between a parent cluster and an attached monomer (A-A' or B-B' bond) slightly stronger $E_{ij}=1.4E$ so that these bonds are harder to be broken by thermal fluctuations. In the process of detachment (SI section II), the monomer bonds become irreversible with $E_{ij}=2.5E$.

\section{IV. Validation of Simulations against Experiment}

The implementation of interaction potential is validated against recent experimental results by Perry \emph{et al.} \cite{perry2015two}. In the experiment, colloidal clusters of 6 particles bound by short-ranged depletion interactions are confined in two spatial dimensions.

In general, the sticky parameter 
\begin{equation}
\kappa = \frac{e^{-\beta U(d)}}{d \sqrt{\frac{2}{\pi} \beta U''(d)}}
\end{equation}
characterizes the short range interactions between particles \cite{holmes2013geometrical}.
In the experiment, the sticky parameter is measured to be $\kappa=30.5$, and we match this value in the simulations by setting the depth of the potential to be $1.107E$, with $E\equiv 10k_BT$, while leaving other parameters and settings of simulation unchanged.

\begin{figure*}
\begin {center}
\includegraphics[width = 17cm]{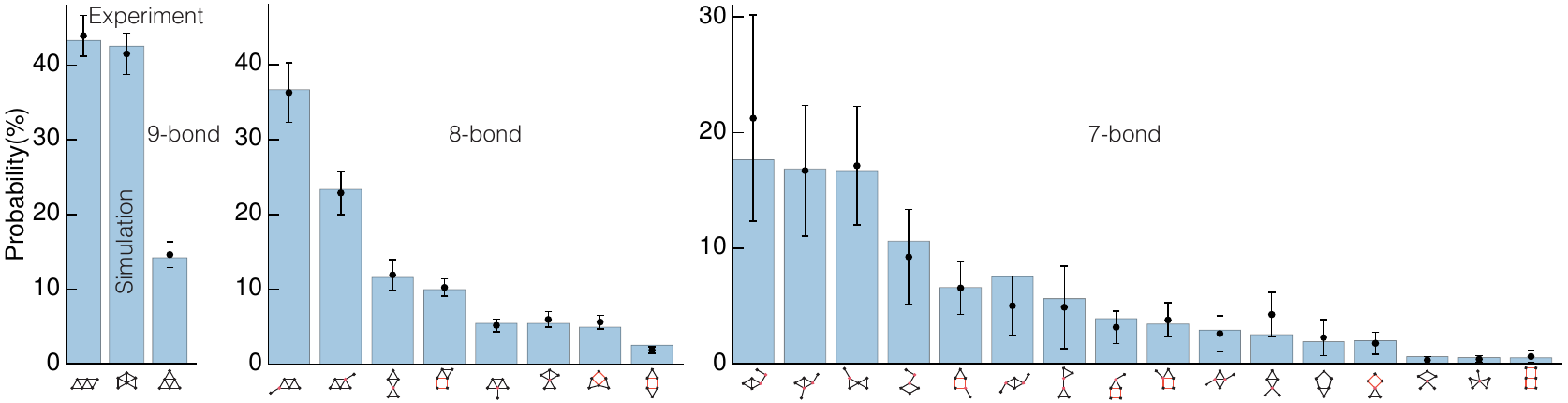} 
\caption{Probability distributions of the 9, 8, and 7 bond states of a colloidal cluster composed of 6 particles in two dimensions. Experimental data (black dots) are from Perry \emph{et al.} \cite{perry2015two}, while results of our validated simulations are the blue bars. Each cluster state is represented by its connectivity graph.
}
\label{figureSI1}
\end{center}
\end{figure*}

To confirm the matching between our simulation and the experiment, we consider the assembly of 6 particle clusters. The 6 particles have three ground states (9 bonds in total) as well as various excited states with 8 and 7 bonds (see Fig.~S\ref{figureSI1}). We ran 20 simulations with duration $10,000(d^2/D)$ each, and counted the different observed cluster states. From this data we constructed the histogram of probabilities of observing the 9, 8 and 7 bond states,
Fig.~S\ref{figureSI1}. The simulation outcomes are in excellent agreement with the experimentally observed probabilities. These results validate the form of the potential that we used and confirm that bond strengths of order $E$ are representative of the experimental situation.

\section{V. Replication pathways}
\begin{figure*}
\begin {center}
\includegraphics[width = 17cm]{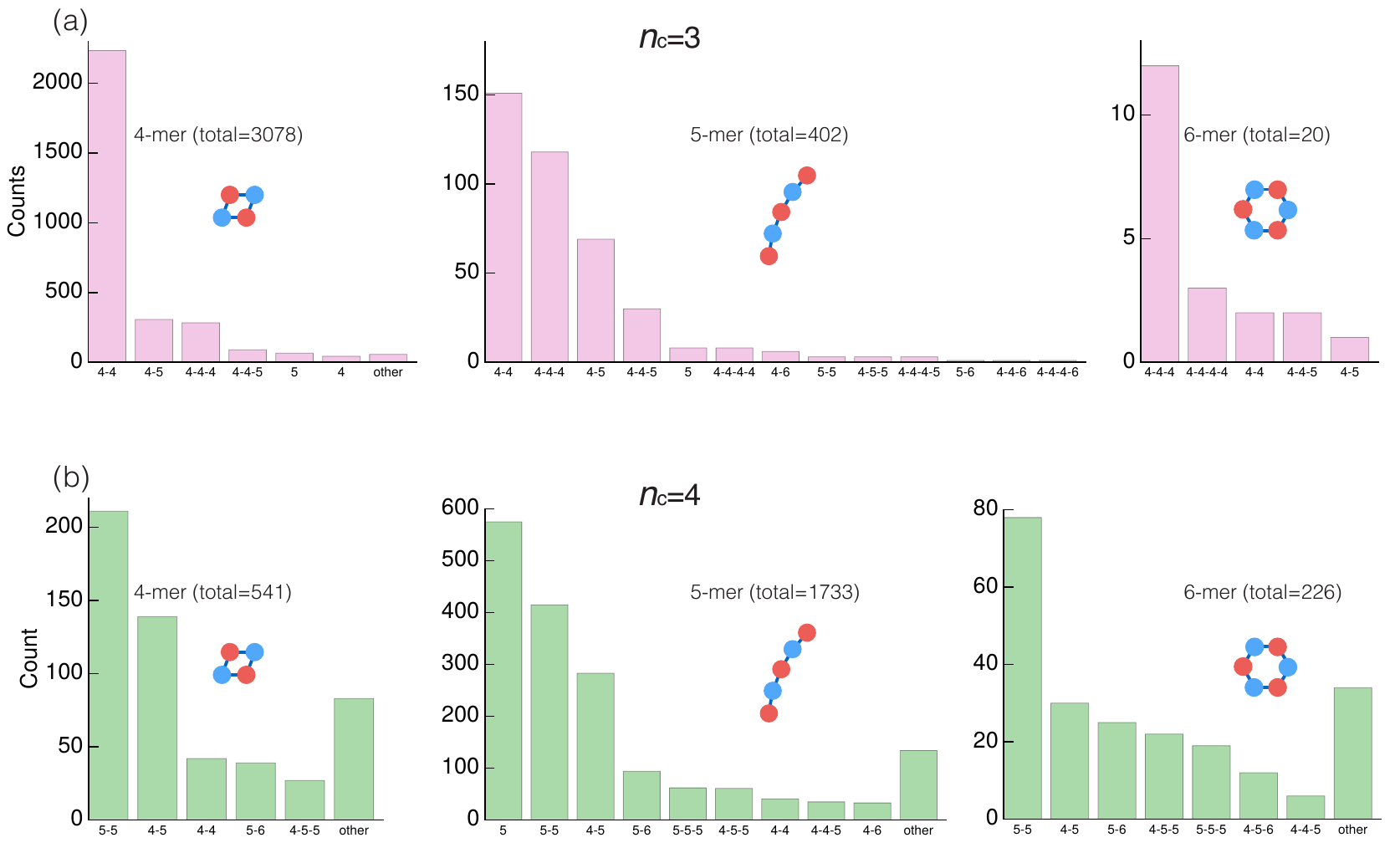} 
\caption{
The number of instances a replication pathway is observed, cumulative from 10 independent simulations. The label on the horizontal axis denotes the sizes of parents which created the offspring depicted in the histogram. Note that the replication instances we observed involved various numbers of parents, between 1 and 4. (a) Replication rule $n_{\rm{c}}=3$. The most frequent replication pathway is ``4-4'', in which two square clusters template another square cluster as described in Fig.~2(b) of the main text. (b) Replication rule $n_{\rm{c}}=4$. The most frequent replication pathway is ``5'', in which single 5-mer parent cluster templates another 5-mer.}
\label{figureSI2}
\end{center}
\end{figure*}

We analyzed the production of various cluster in 10 independent simulations, each with $N=1652$ and $\phi=0.193$. With the rule $n_{\rm{c}}=3$, the simulations cumulatively produced: 3078 of 4-mer clusters, 402 of 5-mer clusters and 20 of 6-mer clusters. Fig.~S\ref{figureSI2} summarizes the frequencies of replication pathways for each cluster geometry.
As expected, the most frequent pathway is the one in which two 4-mer parent clusters template a 4-mer daughter cluster (``4-4'') as shown in Fig.~S\ref{figureSI2}(a). The replication pathways for 5-mers and 6-mers shown in Fig.~2(b) of the main text are the most frequent pathways.

With the mutated rule $n_{\rm{c}}=4$, the simulations cumulatively produced: 541 of 4-mer clusters, 1733 of 5-mer clusters, 226 of 6-mer clusters, 38 of 7-mer clusters and 3 of 8-mer clusters. Fig.~S\ref{figureSI2}(b) shows the frequency of replication pathways for the 4-mer, 5-mer and 6-mer geometries.  Note that self-replication of single 5-mer chain (``5''), which was suppressed under $n_{\rm{c}}=3$ rule is the most frequent pathway under the mutated rule $n_{\rm{c}}=4$.

\section{VI. Effective Diffusion Constant of Parent Clusters}

\begin{figure*}
\begin {center}
\includegraphics[width = 17cm]{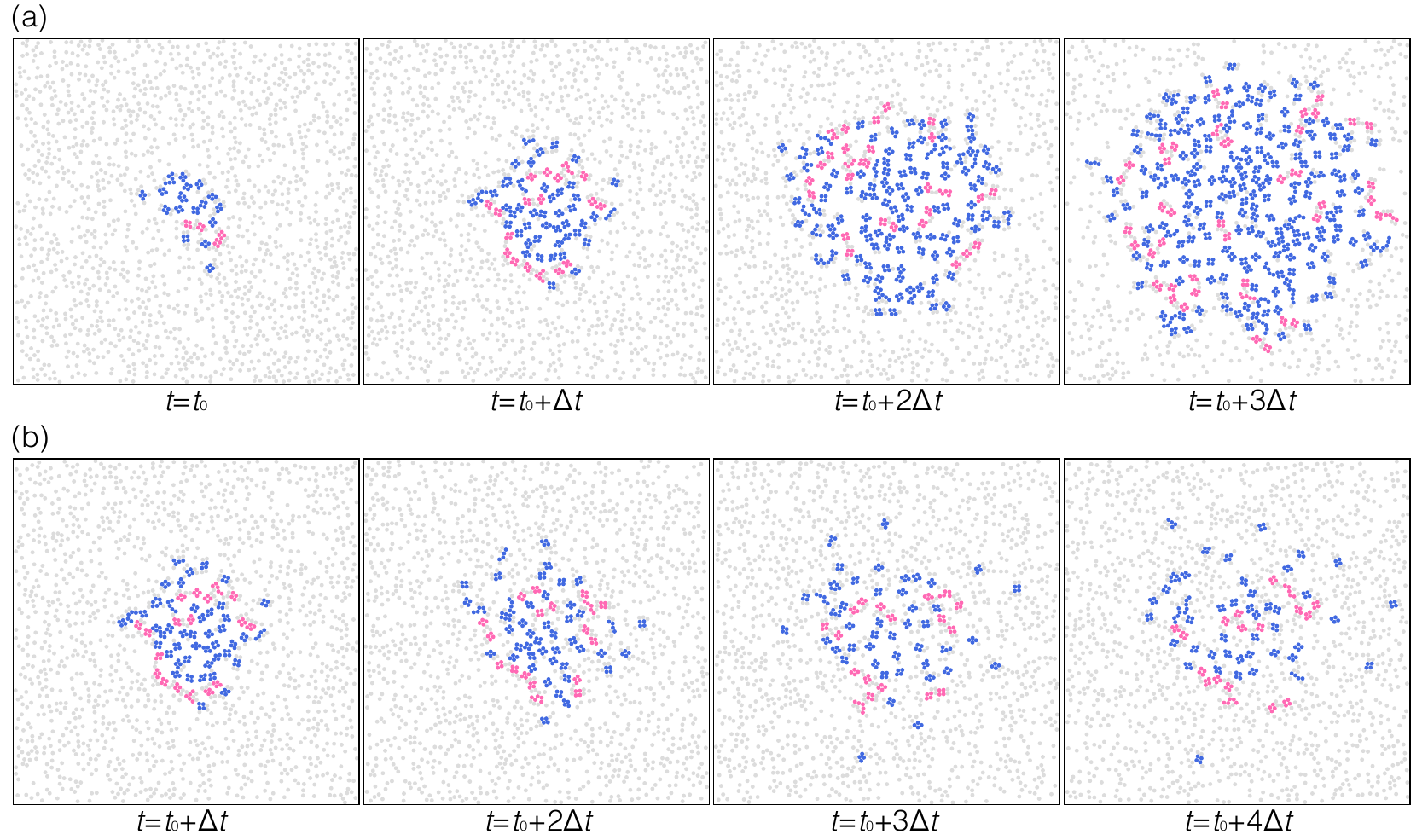} 
\caption{
(a) Snapshots from the simulation in Fig.~3 of the main text with a different coloring: a cluster that is connected with any other cluster via attached monomers is colored in pink, otherwise in dark blue.
(b) Snapshots from direct measurement of $D_{\rm{eff}}^{\rm{sim}}$. We froze all the bonds found at the time $t=t+\Delta t$ (second panel of (a)), preventing formation of new bonds and replication reaction. Thus, clusters diffuse while retaining their attached monomers and connections with other clusters. 
}
\label{figureSI3}
\end{center}
\end{figure*}

When we model the dynamics of front propagation with F-KPP equation, we assume that parent clusters are diffusing while undergoing self-replication reactions.
However, the diffusive component of the cluster behavior may be significantly influenced --- slowed down or even prohibited --- by the effective attraction which is mediated between the clusters by their attached monomers, e.g., stage III or VI in Fig.~1(a) of main text.
To inspect the diffusive behavior and effective attraction between the clusters, in Fig.~S\ref{figureSI3}(a) we color in pink each parent clusters which is connected to any other cluster via attached monomers, and otherwise color in dark blue. Since the colony is not melded into a continuous structure but instead consists of many separate parts that can diffuse, we confirm that both diffusion and self-replication contribute to the observed front propagation.\\

To directly measure the effective diffusion constant $D^{\rm{sim}}_{\rm{eff}}$ of parent clusters
we freeze all bonds at a time when the front is already formed (second panel of Fig.~3 of the main text) and continue the simulation without allowing formation of new bonds, as shown in Fig.~S\ref{figureSI3}(b). We extract the value of $D^{\rm{sim}}_{\rm{eff}}$ directly from the movement of center of mass of each cluster, and average over all clusters to obtain $D^{\rm{sim}}_{\rm{eff}}=0.12\pm 0.04D$. Note that this diffusion constant is smaller than for free, non-interacting square clusters for which we measured $D_{\rm{free\ sq}}=0.25D$, as expected due to effective attraction.

\section{VII. Front width}
\begin{figure*}
\begin {center}
\includegraphics[width = 0.95\textwidth]{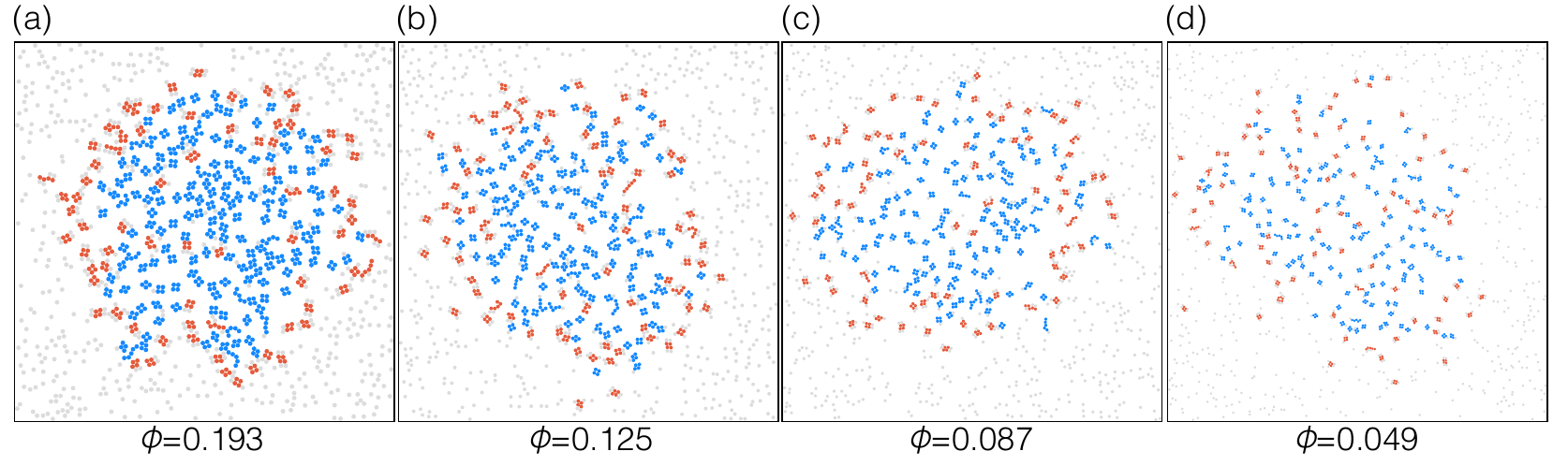}
\caption{(a-d)Snapshots from simulations with decreasing area fraction $\phi$ and fixed $N=1652$. In each panel the time is chosen such that the cluster fill most of the system. Parent clusters with at least two attached monomers are colored red (active parents) and parent clusters with less than two attached monomers are colored blue (inactive parents), see discussion in main text. Note the increasing front width as their fraction $\phi$ is decreased.}
\label{figureSI4}
\end{center}
\end{figure*}

In the Fig.~3 of the main text, the last snapshot shows a fully formed colony. The coloring of active parents, as defined in the main text, indicates well that the width of the front, set by the length scale $\lambda$, is on the order of a few particle diameters. This figure panel is reproduced in Fig.~S\ref{figureSI4}(a). In this case we see that the initial exponential growth of the colony (used to extract growth rate $\alpha$) crosses over with time to a growth consistent with $t^2$. To explore how the formation the front depends on the system parameters we vary the area fraction $\phi$ of the particles. Panels (b-d) of Fig.~S\ref{figureSI4} show snapshots from simulations with progressively smaller $\phi$. In each simulation we choose a long enough time that clusters occupy majority of the system. With smaller $\phi$ the coloring of active parents indicates that $\lambda$ grows. At lowest $\phi=0.049$ we cannot distinguish clearly a front region. In this case $\lambda$ is comparable to the colony size and comparable to the system size $L$. Consistently with these observations, at the lowest area fraction the growth curve remains consistent with exponential growth.

\section{VIII. Survival probabilities for mutated and non-mutated clusters}
Using the 13 simulations, as defined in the main text, we study the survival rates of mutated and non-mutated clusters. In each simulation we trace the progeny of the mutated cluster and the progeny of several randomly selected non-mutated cluster at the opposite side of the colony front. The Fig.~S5 demonstrates the types of outcomes we distinguish: The progeny can form an angular sector (``survival'' outcome), can stop reproducing behind the front (``extinction''), or have a shrinking sector with just a few progeny left on the front (``undecidable''). Using the number of undecidable cases as the estimate for error in the outcome, we find the survival rates quoted in main text.

To understand the origin of difference between survival rates of non-mutated and mutated clusters, we compare simulations of purely non-mutated clusters (only $n_{\rm{c}}=3$) to simulations of purely mutated clusters (only $n_{\rm{c}}=4$). We average over 10 simulations in each case. To quantify the difference between the mutated an non-mutated populations, Fig.~S\ref{figureSI6}(a) shows population of 4-mers, 5-mers, 6-mers, and 7-mers as a function of time in a system where all clusters obey the mutated rule $n_{\rm{c}}=4$. In contrast to the $n_{\rm{c}}=3$ simulations (Fig.~2(c) of main text), the 5-mer cluster (chain geometry) dominates the colony.
The capacity of the populations to consume monomers however does not vary significantly: Fig.~S\ref{figureSI6}(b) shows that the number of monomers consumed is very similar between squares under the $n_{\rm{c}}=3$ rule and 5-mer chains under the $n_{\rm{c}}=4$ rule.
The difference in fitness instead can be attributed to the different growth rates of mutated and non-mutated populations. Fig.~S\ref{figureSI6}(c) shows the total population of all clusters as a function of time for simulations with two different replication rules. With mutated rule the growth rate of the total population is smaller than with the non-mutated rule.

\begin{figure*}
\begin {center}
\includegraphics[width = 17cm]{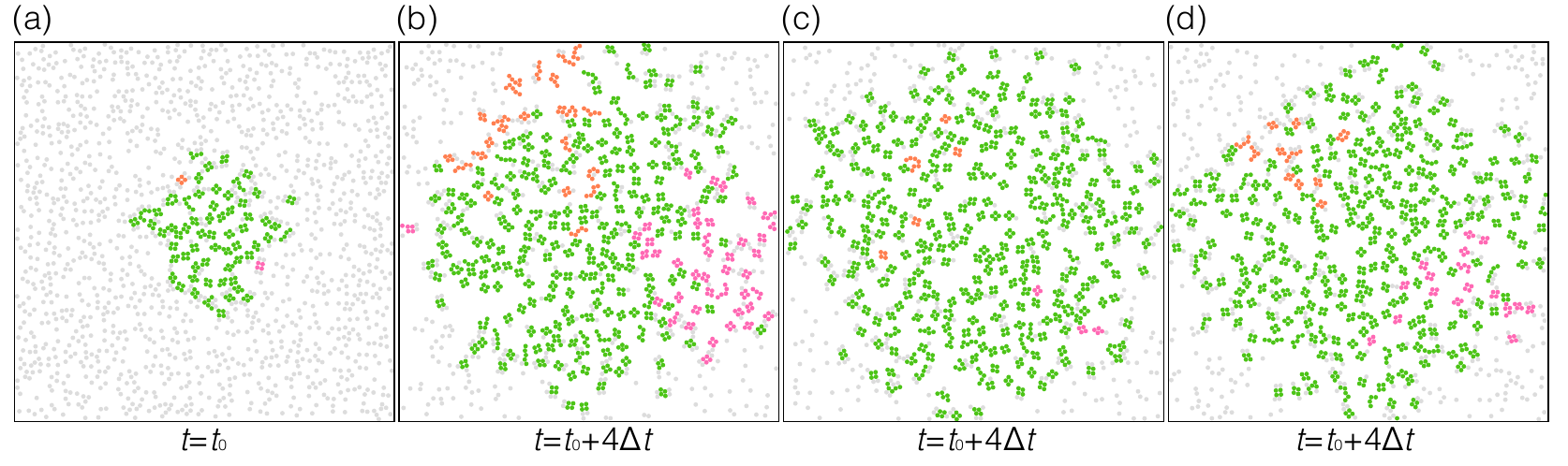} 
\caption{Types of outcomes of a lineage of a mutated cluster (orange) and a
  non-mutated cluster (pink). (a)  The second snapshot ($t=t_{0}+\Delta t$) of Fig.~3 is shown, where one mutated square cluster (orange) and non-mutated cluster (pink) are placed at the boundary to track their lineages. (b-d) Outcomes of independent simulations started from this initial condition. (b) The progeny forms an angular sector and occupies expanding front, which   we classify as ``survival''.  (c) The progeny stops reproducing  behind the front, which we classify as ``extinction''. (d) The progeny has a shrinking sector with just a few clusters left on the  front, which we classify as ``undecidable''. }
\label{figureSI5}
\end{center}
\end{figure*}

\begin{figure*}
\begin {center}
\includegraphics[width = 17cm]{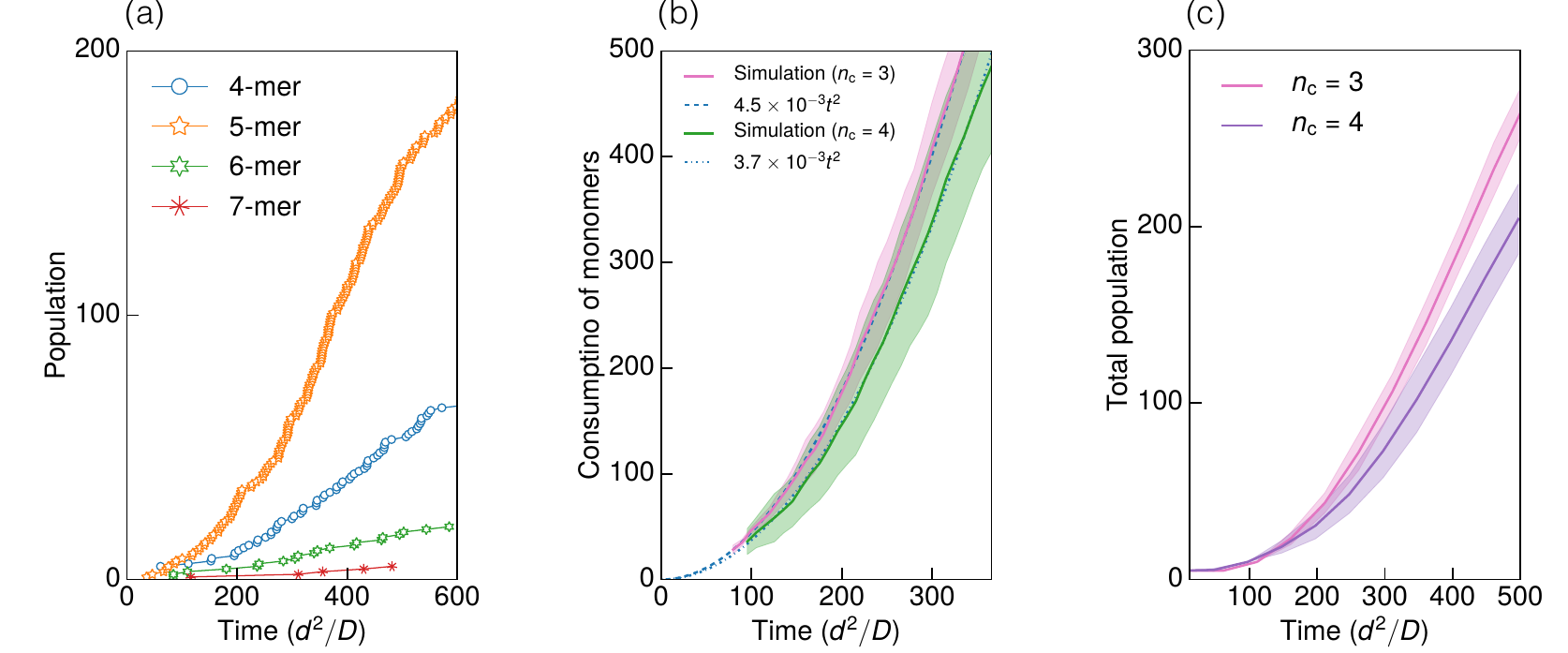} 
\caption{ (a) Population of 4-mers, 5-mers, 6-mers, and 7-mers as a function of time obtained under the mutated replication rule $n_{\rm{c}}=4$. (b) Comparison of the amount of monomers consumed into 5 particle clusters as these are created under $n_{\rm{c}}=4$, to the amount of monomers consumed into square clusters under $n_{\rm{c}}=3$. (c) Comparison of the total populations of clusters as a function of time obtained in simulation with $n_{\rm{c}}=3$ and simulation with $n_{\rm{c}}=4$. The shaded region shows one standard deviation above and below the average.
}

\label{figureSI6}
\end{center}
\end{figure*}

\bibliography{lib,referenceSI}